\documentclass[12pt]{article}
\usepackage{amssymb}

\usepackage{graphicx}
\usepackage{amsmath}
\usepackage{float}
\usepackage[dvips]{color}
\usepackage{rotating}

\renewcommand{\bf}[1]{\textbf{#1}}

\input{tcilatex}

\begin{document}

\begin{titlepage}
\bigskip \begin{flushright}
hep-th/0508200\\
\end{flushright}
\vspace{1cm}
\begin{center}
{\Large \bf {Eguchi-Hanson Solitons in Odd Dimensions}}\\
\end{center}
\vspace{2cm}
\begin{center}
R. Clarkson\footnote{EMail:  r2clarks@astro.uwaterloo.ca}, R. B.
Mann\footnote{EMail:  mann@avatar.uwaterloo.ca} \\
Department of Physics, University of Waterloo, \\
Waterloo, Ontario N2L 3G1, Canada\\
\vspace{1cm}
\end{center}

\begin{abstract}
We present a new class of solutions in odd dimensions to Einstein's
equations containing either a positive or negative cosmological constant. These
solutions resemble the even-dimensional Eguchi-Hanson-(A)dS metrics, with the
added feature of having Lorentzian signatures. They are asymptotic 
to (A)dS$_{d+1}/Z_p$.
In the AdS case their energy is negative relative to that of pure AdS.
We present perturbative evidence in 5 dimensions that such metrics are the states of lowest energy in their
asymptotic class, and present a conjecture that this is generally true for all such metrics. 
In the dS case these solutions have a cosmological horizon. We show that their mass
at future infinity is less than that of pure dS.  
\end{abstract}
\end{titlepage}

\section{Introduction}

Taub-NUT metrics are playing an increasingly important role in physics.
Originally derived in four dimensions \cite{Taub,NUT}, such solutions were
asymptotically locally flat (ALF) solutions to the Einstein equations, with
the NUT charge behaving somewhat like a magnetic mass. Their asymptotically
locally Euclidean (ALE) counterparts provided a new class of self-dual
gravitational instantons, analogous to the self-dual instantons that appear
in Yang-Mills theory \cite{EguchiHanson,GibHawkCMP}. Incorporation of such
objects into Kaluza-Klein theories led to the discovery of the Kaluza-Klein
monopole, and higher-dimensional generalizations followed shortly
thereafter. More recently it has become clear that ALF solutions have
interesting gravitational thermodynamics, since their Euclidean sections
cannot be everywhere foliated by surfaces of constant (Euclidean) time \cite%
{Hawking,Hunter}. The gravitational entropy of these solutions is not
proportional to event-horizon area \cite{GarfMann}. Generalizations of such
solutions to non-zero cosmological constant have thus provided interesting
tests of the AdS/CFT and dS/CFT correspondence conjectures \cite%
{Mann1,Mann2,HHP,Lorenzo}.

Other ALE instantons have also been found, the simplest nontrivial example
perhaps being the Eguchi-Hanson (EH) metric \cite{EguchiHanson}. Both
Taub-NUT and EH metrics are special cases of Atiyah-Hitchin metrics, and it
has been shown in a recent set of papers \cite{branepapers} how all three
types of geometries can be embedded in $M$-theory, providing new M2- and
M5-branes solutions whose distinguishing feature is that the solution is not
restricted to be in the near core region of a D6 (or D5 or D4)-brane. Such
solutions can be compactified on a circle, yielding new solutions of
type-IIA string theory.

Recently it has been shown \cite{EHshort} that generalizations of the EH
metric exist in 5-dimensions that have a number of interesting features.
Referred to as Eguchi-Hanson solitons, they are asymptotic to AdS$_{n}/Z_{p}$
where $p\geq 3$. Unlike the four dimensional case, a Lorentzian signature is
possible, yielding a non-simply connected background manifold for the CFT
boundary theory. They are obtained from a recently-derived 5-dimensional
generalization of the Taub-NUT metric \cite{MannStelea} in a manner
analogous to that used in deriving the AdS soliton. Their spatial sections
approach that of the EH metric for large cosmological constant. In five
dimensions the total energy of these solitons is negative, though bounded
from below consistent with earlier arguments \cite{HoroJac}.

Motivated by the above, in this paper we explore these generalizations of
the EH metric in higher dimensions. Specifically, we obtain a new set of
solutions to the Einstein equations of motion with a cosmological constant,
that provided a natural (odd-dimensional) generalization of the EH metric.
They can be derived from a set of inhomogeneous Einstein metrics on sphere
bundles fibred over Einstein-Kahler spaces that were recently obtained \cite%
{MannStelea,LuPagePope}. We show that in the limit the cosmological constant
vanishes we obtain a set of $d$-dimensional generalizations of the ALE EH
metric. As such, our approach provides a $d$-dimensional generalization for
obtaining the EH metric from a limiting procedure in the 4-dimensional case%
\footnote{%
For an alternate derivation of the EH metric in four dimensions, see ref. %
\cite{Mahapatra}}. We explore some of the implications of the simplest
(5-dimensional) solution for the AdS/CFT correspondence when the
cosmological constant is negative. Unlike the four dimensional case, a
Lorentzian signature is possible, and so an interesting background metric
for the CFT boundary theory is possible. We find that such solutions --
though of negative energy -- are perturbatively stable, and we conjecture
that they are the states of lowest energy in their asymptotic class. In the
dS case we show that these solutions satisfy the maximal mass conjecture.

The outline of our paper is as follows. We begin with the five dimensional
case, and then show how to obtain the solution in odd $d+1$ dimensions from
the aforementioned set of inhomogeneous Einstein metrics on sphere bundles
fibred over Einstein-Kahler spaces. We then discuss the regularity
conditions for the solitons for both types of asymptotia (AdS and dS). In
section 4 we study aspects of the (A)dS/CFT correspondence for these
solitons. We show that in the AdS case they have negative energy but are
perturbatively stable, whereas in the dS case they satisfy the maximal mass
conjecture.

\section{Deriving the Metric}

To illustrate our approach, we review the 5-dimensional case. The metric
here can be derived from the five-dimensional generalization \cite%
{MannStelea} of the Taub-NUT metric, 
\begin{equation}
ds^{2}=-\rho ^{2}dt^{2}+4n^{2}F(\rho )\left[ d\psi +\cos (\theta )d\phi %
\right] ^{2}+\frac{d\rho ^{2}}{F(\rho )}+(\rho ^{2}-n^{2})(d\theta ^{2}+\sin
(\theta )^{2}d\phi ^{2})  \label{5dTN}
\end{equation}%
where the $U(1)$-fibration is a partial fibration over a two-dimensional
subspace the three dimensional base space. The function $F(\rho )$ is 
\begin{equation}
F(\rho )=\frac{\rho ^{4}+4m\ell ^{2}-2n^{2}\rho ^{2}}{\ell ^{2}(\rho
^{2}-n^{2})}  \label{f5dTN}
\end{equation}%
and the condition for this to satisfy the 5D Einstein equations with
cosmological constant $\Lambda =-\frac{6}{\ell ^{2}}$ is $n={\textstyle\frac{%
\ell }{2}}$.

The spacetime is not trivial in the sense that setting the NUT charge
(equivalently, the cosmological constant $\Lambda $) to zero yields a
degenerate metric. We therefore seek a set of transformations that render a
non-trivial metric in the $\Lambda \rightarrow 0$ (or $l\rightarrow \infty $%
) limit. These can be obtained through the following transformations 
\begin{equation}
\rho ^{2}=r^{2}+n^{2}\text{ \ \ \ \ \ }m=\frac{\ell ^{2}}{64}-\frac{a^{4}}{
64\ell ^{2}}  \label{5dTNCHcoords}
\end{equation}
after which we set $r\rightarrow r/2$, $t\rightarrow 2 t/\ell $, thereby
obtaining 
\begin{eqnarray}
ds^{2} &=&-g(r)dt^{2}+\frac{r^{2}f(r)}{4}\left[ d\psi +\cos (\theta )d\phi %
\right] ^{2}+\frac{dr^{2}}{f(r)g(r)}  \notag \\
&&+\frac{r^{2}}{4}(d\theta ^{2}+\sin (\theta )^{2}d\phi ^{2})
\label{mtrcEHdS5} \\
g(r) &=& 1 + \frac{r^{2}}{\ell ^{2}}~~~,~~~~f(r)=1-\frac{a^{4}}{r^{4}} 
\notag
\end{eqnarray}
Note that (\ref{mtrcEHdS5}) solves Einstein's equations with negative
cosmological constant $\Lambda = -6/\ell^2$. Analytically continuing $\ell
\rightarrow \text{i} \ell $ will turn (\ref{mtrcEHdS5}) into a metric
solving Einstein's equation with a positive cosmological constant.

The metric (\ref{5dTN}) -- transformed into the metric (\ref{mtrcEHdS5}) --
provides us with a new means of obtaining the Eguchi-Hanson metric in
4-dimensions. We see that in the $\ell \rightarrow \infty $ limit the metric
(\ref{mtrcEHdS5}) yields the Eguchi-Hanson metric 
\begin{equation}
ds^{2}=\frac{r^{2}}{4}f(r)\left[ d\psi +\cos (\theta )d\phi \right] ^{2}+ 
\frac{dr^{2}}{f(r)}+\frac{r^{2}}{4}(d\theta ^{2}+\sin (\theta )^{2}d\phi
^{2})  \label{4dEH}
\end{equation}
as a $t=$constant hypersurface. Note that the transformations (\ref%
{5dTNCHcoords}) are crucial in obtaining this result; the $\ell \rightarrow
\infty $ limit of (\ref{5dTN}) yields a degenerate metric.

The metric (\ref{mtrcEHdS5}) solves the Einstein equations with a negative
(positive) cosmological constant $\Lambda =\mp {\frac{6}{\ell ^{2}}}$ (or
the vaccum equations when $\ell \rightarrow \infty $). We call these metrics
Eguchi-Hanson-AdS (EHAdS)/Eguchi-Hanson-dS (EHdS) solitons respectively,
since they bear an interesting resemblance to the Eguchi-Hanson metric in
four dimensions. However unlike the four dimensional case, a Lorentzian
signature is possible.

This metric (\ref{mtrcEHdS5}) suggests a generalization to any odd dimension 
$(d+1)$ greater than five. Setting $d=2k+2$, we have found that the
following set of metrics 
\begin{eqnarray}
ds^{2} &=&-g(r)dt^{2}+\left( \frac{2r}{d}\right) ^{2}f(r)\left[ d\psi
+\sum_{i=1}^{k}\cos (\theta _{i})d\phi _{i}\right] ^{2}  \notag \\
&&+\frac{dr^{2}}{g(r)f(r)}+\frac{r^{2}}{d}\sum_{i=1}^{k}d\Sigma _{2(i)}^{2}
\label{EH d-dim}
\end{eqnarray}%
where 
\begin{equation}
d\Sigma _{2(i)}^{2}=d\theta _{i}^{2}+\sin ^{2}(\theta _{i})d\phi _{i}^{2}
\label{dOmegasq}
\end{equation}%
and the metric functions are given by 
\begin{equation}
g(r)=1\mp \frac{r^{2}}{\ell ^{2}}~~~,~~~~~f(r)=1-\left( \frac{a}{r}\right)
^{d}  \label{d-dimmetricfns}
\end{equation}%
satisfy the $(d+1)$-dimensional Einstein equations for both a positive and
negative cosmological constant $\Lambda =\pm d(d-1)/(2\ell ^{2})$.

The $\ell \rightarrow \infty $ limit of this general metric yields 
\begin{equation}
ds^{2}=\left( \frac{2r}{d}\right) ^{2}\left( 1-\left( \frac{a}{r}\right)
^{d}\right) \left[ d\psi +\sum_{i=1}^{k}\cos (\theta _{i})d\phi _{i}\right]
^{2}+\frac{dr^{2}}{1-\left( \frac{a}{r}\right) ^{d}}+\frac{r^{2}}{d}
\sum_{i=1}^{k}d\Sigma _{2(i)}^{2}  \label{ddimEH}
\end{equation}
as a $t=$constant hypersurface of the metric (\ref{EH d-dim}). These can be
regarded as $d$-dimensional generalizations of the Eguchi-Hanson metric (\ref%
{4dEH}).

We now demonstrate how the class of metrics (\ref{EH d-dim}) can be derived
from transforming a class of Einstein metrics on $S^{m+2}$ sphere bundles
fibred over $2n$-dimensional Einstein-K\"{a}hler spaces \cite{LuPagePope} 
\begin{equation}
ds^{2}=\frac{(1-\rho ^{2})^{n}d\rho ^{2}}{P(\rho )}+\frac{c^{2}P(\rho )}{
(1-\rho ^{2})^{n}}\left[ d\tau -2A\right] ^{2}+c(1-\rho ^{2})d\Sigma
_{2n}^{2}+\frac{(m-1)\rho ^{2}}{\Lambda -\lambda /c}d\Omega _{m}^{2}
\label{mtrclpp}
\end{equation}
with $(d+1)=2n+m+2$, and $A$ is a 1-form potential for the K\"{a}hler form $
d\Sigma _{2n}^{2}$ ($J=dA$). The function $P(r)$ obeys the differential
equation 
\begin{equation}
\frac{d}{d{\rho }}\left[ \rho ^{m-1}P(\rho )\right] =\rho ^{m-2}\left[
\Lambda \left( 1-\rho ^{2}\right) ^{n+1}-\frac{\lambda }{c}\left( 1-\rho
^{2}\right) ^{n}\right]  \label{Prdq}
\end{equation}
whose given in terms of a linear combination of hypergeometric functions 
\begin{eqnarray}
P(\rho ) &=&\frac{\Lambda }{m-1}~~{}_{2}F_{1}\left( -n-1,\frac{m-1}{2};\frac{
m+1}{2};\rho ^{2}\right)  \notag \\
&&-\frac{\lambda }{c(m-1)}~~{}_{2}F_{1}\left( -n,\frac{m-1}{2};\frac{m+1}{2}
;\rho ^{2}\right) +\mu \rho ^{1-m}  \label{Pr}
\end{eqnarray}

Carrying out the higher-dimensional generalization of the transformation (%
\ref{5dTNCHcoords}), we write $\rho^{2}=1+x$, which renders (\ref{Prdq}) as 
\begin{equation}
\frac{d}{d{x}}F(x)=(-1)^{n+1}\frac{\Lambda }{2}\left( 1+x\right) ^{(m-3)/2} %
\left[ x^{n+1}+\nu x^{n}\right]
\end{equation}
where we have set $\frac{\lambda }{c}=\nu \Lambda $, with $\nu $ an
arbitrary parameter (note \cite{LuPagePope} used $\frac{\lambda }{c}=\frac{%
\Lambda }{\nu }$). This equation has the solution 
\begin{eqnarray}
F(x) &=&\frac{(-1)^{n+1}\Lambda }{2(n+2)}\left\{ x^{n+2}{}_{2}F_{1}\left( %
\left[ -\frac{(m-3)}{2},n+2\right] ,n+3,-x\right) \right.  \label{Fr} \\
&&+\left. \nu \frac{(n+2)}{(n+1)}x^{n+1}{}_{2}F_{1}\left( \left[ -\frac{
(m-3) }{2},n+1\right] ,n+2,-x\right) \right\} +\mu  \notag
\end{eqnarray}
Now taking $x=r^{2}/l^{2}$ and $\tau =2\psi $ in the metric (\ref{mtrclpp})
yields 
\begin{eqnarray}
ds^{2} &=&\mathcal{A}\Bigg[\frac{(m-1)}{\Lambda (1-\nu )}g(r)d\Omega
_{m}^{2}+\frac{(-1)^{n}4c^{2}r^{2}\tilde{f}(r)}{l^{4}g(r)^{(m-1)/2}} \left[
d\psi -A\right] ^{2}  \notag \\
&&+\frac{(-1)^{n}g(r)^{(m-3)/2}dr^{2}}{\tilde{f}(r)}-\frac{cr^{2}}{l^{2}}
d\Sigma _{2n}^{2}\Bigg]  \label{mtrcgen}
\end{eqnarray}
where 
\begin{eqnarray}
\mathcal{A} &=&\frac{2n+m}{2}  \notag \\
g(\rho ) &=&1+\frac{r^{2}}{l^{2}}  \label{mtrcgenfns} \\
c &=&\frac{1}{\Lambda \nu }=\frac{2 l^{2}}{d(d-1)\nu }  \notag
\end{eqnarray}
and 
\begin{eqnarray}
\tilde{f}(r) &=&\frac{l^{2}F(x)}{x^{n+1}}  \notag \\
&=&\frac{(-1)^{n+1}\Lambda l^{2}}{2(n+2)}\left\{ \frac{r^{2}}{l^{2}}
~{}_{2}F_{1}\left( \left[ -\frac{(m-3)}{2},n+2\right] ,n+3,-\frac{r^{2}}{
l^{2}}\right) \right.  \label{Fgen} \\
&&\left. +\frac{\nu (n+2)}{(n+1)}~{}_{2}F_{1}\left( \left[ -\frac{(m-3)}{2}
,n+1\right] ,n+2,-\frac{r^{2}}{l^{2}}\right) \right\} +\frac{\mu l ^{2n+4}}{%
r^{2n+2}}  \notag
\end{eqnarray}

There are two free parameters in the class of metrics (\ref{mtrcgen}): $\nu $
and $\mu $; the quantity $l$ simply represents a length scale for the
metric. Note that we cannot set $l\rightarrow \infty $ in (\ref{mtrcgen}) as
written (but see below). We have explicitly checked (\ref{mtrcgen}), with ( %
\ref{mtrcgenfns}) and (\ref{Fgen}), solve Einstein's equations for $m=2,3,4$
and $n=1,2,3,4$.

We can now derive our general dimensional EH--type metrics (\ref{EH d-dim})
from this more general metric (\ref{mtrcgen}), by taking $m=1$ (which will
require $\nu = 1$). This first requires that we redefine the coordinate we
will get from the $d\Omega_{m=1}$ term to get rid of divergences; i.e.
transform $\varphi = \frac{\Lambda (1-\nu)}{(m-1)} t$, to give 
\begin{equation*}
\frac{(m-1)}{\Lambda (1-\nu)} d\Omega_1 = \frac{(m-1)}{\Lambda (1-\nu)}
d\varphi = dt
\end{equation*}
before taking $m=1$ in the coefficient. Note that there is no problem taking 
$m=1$, $\nu=1$ in any of the other terms in (\ref{mtrcgen}), nor is there a
problem in (\ref{Fgen}).

Thus, comparing to the general--dimensional case, we see that the quantity $
d\Omega _{m}^{2}$ plays the role of time, and with appropriate analytic
continuation we can choose the signature accordingly. The generalized EH
metrics (\ref{EH d-dim}) are obtained after several steps by first setting $
\nu = m = 1$ with 
\begin{equation*}
\mu =-\frac{(-1)^{n+1} \Lambda b^{2n+2}}{(2n+2) l^{2n+2}}
\end{equation*}%
and $\Lambda = 1/l^2, c = l^2$, $\mathcal{A} = 1$, recalling that $d = 2n+2$
when $m=1$. Then, using the further rescalings 
\begin{equation*}
r=\frac{r}{\sqrt{d}},~b=\frac{a}{\sqrt{d}} \text{ \ \ \ and \ \ \ } l=\frac{%
\ell}{\sqrt{d}}
\end{equation*}
one gets the general metric (\ref{EH d-dim}). Note that the Eguchi--Hanson
metrics (\ref{ddimEH}) can then be obtained in the $\ell \rightarrow \infty$
limit.

For $m>1$, we can take the $\Lambda \rightarrow 0$ limit in (\ref{mtrcgen})
by redefining the first coordinate of the $d\Omega _{m}$ sub-metric to
absorb $\Lambda $, i.e. $\varphi =\sqrt{\Lambda }\sigma $, so that as $%
\Lambda \rightarrow 0$, 
\begin{eqnarray}
\frac{1}{\Lambda }d\Omega _{m}^{2} &=&\frac{1}{\Lambda }\left( d\varphi
^{2}+\sin ^{2}(\varphi )d\Omega _{m-1}^{2}\right) =d\sigma ^{2}+\frac{\sin
^{2}(\sqrt{\Lambda }\sigma )}{\Lambda }d\Omega _{m-1}^{2}  \notag \\
&\rightarrow &d\sigma ^{2}+\sigma ^{2}d\Omega _{m-1}^{2}  \label{dOmL0}
\end{eqnarray}%
Thus, the ALF version of (\ref{mtrcgen}) is given by 
\begin{eqnarray}
ds^{2} &=&\frac{(m-1)}{(1-\nu )}\left( d\sigma ^{2}+\sigma ^{2}d\Omega
_{m-1}^{2}\right) +\frac{(-1)^n dr^{2}}{\tilde{h}(r)} + \frac{(-1)^n r^{2}}{%
2(n+1)}d\Sigma_{2n}^{2}  \notag \\
&& + (-1)^n \left( \frac{2r}{2n+2} \right)^{2} \tilde{h}(r)\left[ d\psi -A%
\right] ^{2}  \label{mtrcgenflat} \\
\tilde{h}(r) &=&1-\left( \frac{a}{r}\right) ^{d}  \notag
\end{eqnarray}%
where we have used the following rescalings; 
\begin{eqnarray*}
c & = & \frac{2l^2}{d(d-1)\nu} \\
r & \rightarrow & \sqrt{C(\nu)} r \\
b & = & \left(C (\nu) \right)^{(2n+4)/(4n+4)} a
\end{eqnarray*}
with 
\begin{equation*}
C(\nu) = \frac{(-1)^{n+1} d(d-1) \nu}{4(n+1)}
\end{equation*}
and thus 
\begin{equation}
\tilde{f} (r) = C(\nu) \left[ 1 + \left( \frac{a}{r} \right)^{2n+2} \right]
= C(\nu) \tilde{h} (r)  \label{Fgenflat}
\end{equation}

We have explicitly checked that (\ref{mtrcgenflat}), with (\ref{Fgenflat}),
solves the Einstein equations for $n=1,2,3,4~$and $~m=2,3,4$.

We pause here to comment on the situation in the even-dimensional case. We
find that this case can be obtained as a special case of a class of metrics
derived by Gauntlett \textit{et. al.} \cite{Gauntletetal} that are very
similar to the metrics (\ref{mtrclpp}) discussed above. They have the form
of the (even--dimensional) Eguchi--Hanson--AdS metrics (\ref{ddimEH}), with
a metric 
\begin{eqnarray*}
ds^{2} &=&\frac{r^{2}f(r)}{4}\left[ d\psi +A\right] ^{2}+\frac{dr^{2}}{f(r)}+%
\frac{r^{2}}{4}d\Sigma _{(d-1)}^{2} \\
f(r) &=&\frac{4k}{n}-\left( \frac{a}{r}\right) ^{n}+\frac{(n-1)r^{2}}{
(n+2)\ell ^{2}}
\end{eqnarray*}%
where $k=\left\{ 1,0,-1\right\} $ for the normalized curvature of the base
space, and $n=d+1$ here. For example, in four dimensions, we have 
\begin{equation*}
\left\{ A,d\Sigma _{2}\right\} =\left\{ 
\begin{array}{lcl|l}
\cos (\theta )d\phi & , & (d\theta ^{2}+\sin ^{2}(\theta )d\phi ^{2}) & k=1%
\text{ (spherical)} \\ 
\theta d\phi & , & (d\theta ^{2}+d\phi ^{2}) & k=0\text{ (toroidal)} \\ 
\cosh (\theta )d\phi & , & (d\theta ^{2}+\sinh ^{2}(\theta )d\phi ^{2}) & 
k=-1\text{ (hyperbolic)}%
\end{array}%
\right\}
\end{equation*}%
Thus, in the limit that the cosmological constant vanishes, the Lu, Page and
Pope metric (\ref{mtrclpp}), our general Eguchi--Hanson metric (\ref{ddimEH}%
) and the Gauntlett \textit{et. al.} metrics all have a similar form.

\section{Metric Regularity}

We consider here a more detailed study of the properties of the metric (\ref%
{mtrcEHdS5}). Its Ricci scalar is easily seen to be free of singularities,
and vanishes in the $\ell \rightarrow \infty $ limit. Its Kretschmann scalar 
$\mathcal{R}_{abcd}\mathcal{R}^{abcd}$ is (in the AdS/dS case) 
\begin{equation}
\mathcal{R}_{abcd}\mathcal{R}^{abcd}=\frac{8\left( 5r^{12}+48\ell
^{4}a^{8}\pm 48\ell ^{2}r^{2}a^{8}+9r^{4}a^{8}\right) }{r^{12}\ell ^{4}}
\end{equation}%
($\pm $ for AdS/dS, respectively) whose singular behaviour as $r\rightarrow
0 $ is avoided since it lies inside $r=a$. It is therefore not part of the
spacetime; hence the metric (\ref{mtrcEHdS5}) is free of scalar curvature
singularities. However string-like singularities can arise at $r=a$, and
must be dealt with separately. These can be eliminated in the usual way.
Consider the the behaviour of the metric (\ref{mtrcEHdS5}) as $r\rightarrow
a $. Regularity in the $\left( r,\psi \right) $ section implies that $\psi $
has period $\frac{2\pi }{\sqrt{g(a)}}$ and elimination of string
singularities at the north and south poles $\left( \theta =0,\pi \right) $
implies that an integer multiple of this quantity must equal $4\pi $. This
implies in the asymptotically AdS case that 
\begin{equation}
a^{2}=\ell ^{2}\left( \frac{p^{2}}{4}-1\right)  \label{EHmatch}
\end{equation}%
where $p$ is an integer with $p\geq 3$, yielding in turn that $a>\ell $. The
metric can be written as 
\begin{equation}
ds^{2}=-\left( 1+\frac{r^{2}}{\ell ^{2}}\right) dt^{2}+\frac{r^{2}}{p^{2}}%
f(r)\left[ d\psi +\frac{p}{2}\cos (\theta )d\phi \right] ^{2}+\frac{dr^{2}}{%
\left( 1+\frac{r^{2}}{\ell ^{2}}\right) f(r)}+\frac{r^{2}}{4}d\Omega
_{2}^{2}~~~
\end{equation}%
where now 
\begin{equation*}
f(r)=1-\frac{\ell ^{4}}{r^{4}}\left( \frac{p^{2}}{4}-1\right) ^{2}
\end{equation*}

The asymptotically dS case must be handled with more care, and we must
consider separately the cases where \ $a^{2}<\ell ^{2}$, $a^{2}>\ell ^{2}$
and $\ a^{2}=\ell ^{2}$. If $a^{2}<\ell ^{2}$, then the regularity condition
now equates $\frac{2\pi }{\sqrt{\left| g(a)\right| }}=\frac{4\pi }{p}$ whose
only solution is $p=1$, yielding $a^{2}=\frac{3}{4}\ell ^{2}$. If $%
a^{2}>\ell ^{2}$ then the metric (\ref{mtrcEHdS5}) has closed timelike
curves (CTCs) in the region $a>r>\ell $, and the coordinates $t$ and $\psi $
interchange roles. There is now a cosmological horizon at $r=a$ and the
regularity condition implies that $a^{2}=\ell ^{2}\left( \frac{p^{2}}{4}%
+1\right) $ with all integer values of $p$ allowed. However $r=\ell $\ is no
longer an horizon, since the metric will change signature at that point.
Consequently the spacetime is only defined for $r\geq \ell $. The metric can
be rendered smooth at $r=\ell $ by requiring $t$ to be a periodic variable,
with period $\frac{2\pi \ell ^{3}}{\sqrt{ a^{4}-\ell ^{4}}}=\frac{8\pi \ell 
}{p\sqrt{p^{2}+8}}$.

The case $a=\ell $ is a special case. The metric is not static for $r>a$,
and the metric function $g_{rr}$ has a double root at $r=a$. Consequently
this location is in the infinite past, and the regularity condition no
longer applies. The spacetime expands from this point to an asymptotically
de Sitter region as $r\rightarrow \infty $.

Analysis for the higher dimensional case (\ref{EH d-dim}) is similar.
Elimination of singularities for a metric of the form 
\begin{equation*}
ds^{2}=F(r)\left[ d\psi +\sum_{i=1}^{k}\cos (\theta _{i})d\phi _{i}\right]
^{2}+\frac{dr^{2}}{G(r)}
\end{equation*}%
is attained by demanding consistency between the removal of conical
singularities in the $\left( r,\psi \right) $ section with the criterion
that $\int \cos (\theta _{i})d\phi _{i}=4\pi $. This means that 
\begin{equation}
\frac{4\pi }{\sqrt{\left| F_{+}^{\prime }G_{+}^{\prime }\right| }}=\frac{%
4\pi }{p}  \label{regularity}
\end{equation}%
where $p$ is an integer, with $r=r_{+}$ is the simultaneous root of $F$ and $%
G$. For the metric (\ref{EH d-dim}) $F_{+}^{\prime }G_{+}^{\prime }=4\left(
1\pm \frac{a^{2}}{\ell ^{2}}\right) $ where $g(r)=\left( 1\pm \frac{r^{2}}{
\ell ^{2}}\right) $, yielding the condition (\ref{EHmatch}) above for $(d+1)$%
-dimensions in the AdS case, and the conditions described in the previous
two paragraphs in the dS case. It is straightforward to check that the
Kretschmann scalar is in general finite, since its only possible singularity
is at $r=0$, a point not within the spacetime.

Generalizations of the metric (\ref{EH d-dim}) in which the base space is a
product of $d_{i}$-dimensional spaces of constant curvature (where $\sum
d_{i}=d-2$) are also possible \cite{CrisRobb}. In this case the right-hand
side of the regularity condition (\ref{regularity}) is modified so that it
is multiplied by the normalized curvature of the smallest-dimensional space
in the product in the base \cite{CrisRobb}; we shall not discuss these
solutions here.

The singularity-free form of the (\ref{EH d-dim}) metric is therefore 
\begin{eqnarray}
ds^{2} &=&-g(r)dt^{2}+\left( \frac{r}{p}\right) ^{2}f(r)\left[ d\psi +\frac{%
2p}{d}\sum_{i=1}^{k}\cos (\theta _{i})d\phi _{i}\right] ^{2}  \notag \\
&&+\frac{dr^{2}}{g(r)f(r)}+\frac{r^{2}}{d}\sum_{i=1}^{k}d\Sigma _{2(i)}^{2}
\label{EHAdSpdreg}
\end{eqnarray}%
with 
\begin{equation*}
g(r)=1\pm \frac{r^{2}}{\ell ^{2}}~~,~~~f(r)=1-\frac{a^{d}}{r^{d}}
\end{equation*}%
In the 5-dimensional case the metric (\ref{mtrcEHdS5}) becomes 
\begin{equation}
ds^{2}=-g(r)dt^{2}+\frac{r^{2}}{p^{2}}f(r)\left[ d\psi +\frac{p}{2}\cos
(\theta )d\phi \right] ^{2}+\frac{dr^{2}}{g(r)f(r)}+\frac{r^{2}}{4}%
d\Omega_{2}^{2}~~~  \label{EH(A)dS5p}
\end{equation}%
which can be considered as the singularity-free EH(A)dS metric in five
dimensions \cite{EHshort}.

The regularity condition (\ref{regularity}) implies that the metrics (\ref%
{EHAdSpdreg}) are asymptotic to AdS$_{d+1}/Z_{p}$ where $p\geq 3$. In this
context the the existence of extra light states in a gauge theory formulated
on a quotient space that can be regarded as the boundary of an
asymptotically AdS spacetime is implied by the AdS/CFT correspondence
conjecture. In 5 dimensions the conjecture states that string theory on
spacetimes that asymptotically approach AdS$_{5}$\ $\times $ $S^{5}$ is
equivalent to a conformal field theory (CFT) ($N=4$\ super Yang-Mills $U(N)$
gauge theory) on its boundary $\left( S^{3}\times \mathbb{R}\right) \times
S^{5}$. Finite size effects on the gravity side can be shown to become
important at high temperatures $T\backsim 1/\ell $ (where $\ell $ is the AdS
radius) \cite{HoroJac}; the gauge theory in this situation is in a thermal
state described by the Schwarzschild-AdS solution. The correspondence
implies that the density of low energy states is not affected even though
the volume of $S^{3}$ has been reduced to $S^{3}/\Gamma $ --hence there must
exist light states of the type described above. Our solutions suggest that
these results will carry over to AdS$_{d+1}$ $\times $ $S^{9-d}$ in string
theory, whenever the $d$-dimensional CFT exists.

If the CFT contains fermions, these must be antiperiodic in $\psi$. This
breaks supersymmetry, but it has been pointed out that ordinary (i.e.,
nonsupersymmetric) Yang-Mills gauge theory may be described by compactifying
one direction on a circle and requiring antiperiodic boundary conditions for
the fermions around it \cite{Witten}. This correponds on the supergravity
side to considering spacetimes that are asymptotically locally AdS. Hence
when fermions are present $p$ must be even because going $p$ times along the 
$\psi$ direction yields a circle that is asymptotically contractible, and
the situation is the same as if the asymptotic space were $S^3$ (and not $%
S^{3}/\Gamma$). For $S^3$ the fermions are periodic, and so $p$ must be even
when fermions are present in the CFT.

While there is no horizon in the EHAdS metric, in the EHdS metric (negative
sign in $g(r)$) there is a cosmological horizon at $r=\ell $. In the limit
as $r\rightarrow \infty $, the metric (\ref{EHAdSpdreg}) takes the form 
\begin{equation}
ds^{2}\approx \mp \frac{r^{2}}{\ell ^{2}}dt^{2}+\frac{r^{2}}{p^{2}}\left[
d\psi +N\cos (\theta )d\phi \right] ^{2}\pm \frac{\ell ^{2}dr^{2}}{r^{2}}+%
\frac{r^{2}}{4}d\Omega _{2}^{2}
\end{equation}%
where the upper (lower) sign is the AdS (dS) case. Note that in the dS case
as expected, there is a sign flip when crossing the $r=\ell $ horizon, so
the signature flips to $(+,+,-,+,+)$. So, as $r\rightarrow \infty $ the
metric is $\mathbb{R}\times \mathbb{S}^{3}$ with a twisted $\mathbb{S}^{3}$
where $r$ plays the role of ``time'' in the dS case.

\section{(A)dS/CFT}

In this section we investigate the thermodynamic properties of the metric (%
\ref{mtrcEHdS5}) using the (A)dS/CFT-inspired counter-term method.

A full description of the method can be found in \cite%
{Mann1,Mann2,AdSCFT,BaladS,GM1}. Briefly, however, the AdS/CFT
correspondence conjecture is a holographic relationship between the bulk AdS
spacetime and the conformal field theory (CFT) on the boundary. The
conjecture posits the relationship 
\begin{eqnarray}
Z_{AdS}[\gamma ,\Psi _{0}] &=&\int_{[\gamma ,\Psi _{0}]}D\left[ g\right] D%
\left[ \Psi \right] e^{-I\left[ g,\Psi \right] }=\left\langle \exp \left(
\int_{\partial \mathcal{M}{_{d}}}d^{d}x\sqrt{g}\mathcal{O}_{[\gamma ,\Psi
_{0}]}\right) \right\rangle  \notag \\
&=&Z_{CFT}[\gamma ,\Psi _{0}]  \label{PAR}
\end{eqnarray}%
between the partition functions of any field theory in AdS$_{d+1}$ and a CFT
on the boundary of this space. This relationship suggests counter terms that
can be added to the divergent gravitational action, so that 
\begin{equation}
I=I_{B}+I_{\partial B}+I_{ct}  \label{Itotgen}
\end{equation}%
where $I_{B}$ and $I_{\partial B}$ are the usual Einstein--Hilbert and
Gibbons--Hawking actions, respectively. The counter term action $I_{ct}$,
depending only on quantities intrinsic to the boundary and hence leaving the
equations of motion unchanged, serves to cancel the divergences of the first
two actions. From this total action, one can compute a finite action for a
spacetime, without having to reference a background metric that will
subtract out divergences. Also, upon variation of this full action (\ref%
{Itotgen}), the stress-tensor can be obtained, from which finite conserved
charges can be calculated using the relationship 
\begin{equation}
\mathfrak{D}_{\xi }=\oint_{\Sigma }d^{d-1}S^{a}\xi ^{b}T_{ab}^{\text{eff}}
\label{Qconsgen}
\end{equation}%
If $\xi $ is the time-like killing vector, then one can calculate the
conserved mass; if it is a rotational killing vector, one gets the conserved
angular momentum of the spacetime.Full expansions for the counter term
action and stress-tensor can be found in \cite{Mann1,Mann2,AdSCFT,BaladS,GM1}
and references therein.

\subsection{Comparison of AdS to field theory}

In the asymptotically locally AdS case, there is no horizon. We find from
the counter-term method that the conserved mass is 
\begin{equation}
\mathfrak{M}=\frac{\pi (3\ell ^{4}-4a^{4})}{32G\ell ^{2}p}  \label{Mass5d}
\end{equation}%
and the action is 
\begin{equation}
I_{tot}=\frac{\beta \pi (-8r_{+}^{4}+4a^{4}+3\ell ^{4})}{32\ell ^{2}p}
\label{Itot}
\end{equation}%
where the period $\beta =1/T$ can be chosen arbitrarily since there is no
horizon.

Applying the Gibbs-Duhem relation \cite{GM1} $S=\beta M-I$, the entropy is 
\begin{equation}
S_{tot}=\frac{\beta \pi (r_{+}^{4}-a^{4})}{4\ell ^{2}p}  \label{Stot}
\end{equation}%
Substituting in $r_{+}=a$ in the EH-AdS case gives $S_{tot}=0$, as would be
expected with no horizon.

Similar to the procedure done in \cite{AwadJohnson1}, we can compare our
result (\ref{Mass5d}) with those of the field theory on the boundary of the
AdS$_{5}$ orbifold. Since the local geometry is unchanged, except for the
volume of the $S^{3}$ becoming that of $S^{3}/\Gamma $, the calculation here
will proceed as in the AdS$_{5}$ case \cite{AwadJohnson1}. The stress-tensor
from the gravity side can be computed, including the counter-terms, and can
be shown to have the components 
\begin{eqnarray}
8\pi GT_{tt} &=& -\left\{ \frac{3\ell ^{4}-4a^{4}}{8\ell ^{3}r^{2}}+\frac{%
3\ell ^{4}-4a^{4}}{16\ell r^{4}}\right\} +\mathcal{O}\left( \frac{1}{r^{6}}%
\right)  \notag \\
8\pi GT_{\psi \psi } &=&\frac{12a^{4}-\ell ^{4}}{32\ell r^{2}}+\frac{\ell
(28a^{4}+3\ell ^{4})}{64r^{4}}+\mathcal{O}\left( \frac{1}{r^{6}}\right)
\label{Tgravity} \\
8\pi GT_{\psi \phi } &=&\frac{(12a^{4}-\ell ^{4})\cos (\theta )}{32\ell r^{2}%
}+\frac{(28a^{4}+3\ell ^{4})\ell \cos (\theta )}{64r^{4}}+\mathcal{O}\left( 
\frac{1}{r^{6}}\right)  \notag \\
8\pi GT_{\theta \theta } &=&-\frac{4a^{4}+\ell ^{4}}{32\ell r^{2}}+\frac{%
\ell (3\ell ^{4}-20a^{4})}{64r^{4}}+\mathcal{O}\left( \frac{1}{r^{6}}\right)
\notag \\
8\pi GT_{\phi \phi } &=&\frac{16a^{4}\cos ^{2}(\theta )-\ell ^{4}-4a^{4}}{%
32\ell r^{2}}+\frac{\ell (3\ell ^{4}+48a^{4}\cos ^{2}(\theta )-20a^{4})}{%
64r^{4}}  \notag \\
&&+\mathcal{O}\left( \frac{1}{r^{6}}\right)  \notag
\end{eqnarray}

Removing the factor $r^{2}/\ell ^{2}$ from the metric as $r\rightarrow
\infty $ gives the metric of the conformal field theory, 
\begin{equation}
ds^{2}=-dt^{2}+\frac{\ell ^{2}}{4}\left[ d\psi +\cos (\theta )d\phi \right]
^{2}+\frac{\ell ^{2}}{4}d\Omega _{2}^{2}  \label{mtrcCFT}
\end{equation}%
which has a vanishing Weyl tensor. Since this makes the metric conformally
flat, the expression from Birrell and Davies \cite{BirrellDavies} can be
used for the stress tensor of a field theory on a conformally flat spacetime
in four dimensions: 
\begin{equation}
\left\langle \hat{T}_{ab}^{s}\right\rangle =-\frac{1}{16\pi ^{2}}\left[ 
\frac{\alpha ^{s}}{9}H_{ab}^{(1)}+2\beta ^{s}H_{ab}^{(3)}\right]
\label{Tab4d}
\end{equation}%
where the $\alpha ^{s},\beta ^{s}$ and $n^{s}$ are 
\begin{equation}
\begin{array}{c|c|c|c|}
s & n^{s} & \alpha ^{s} & \beta ^{s} \\ \hline
0 & 6N_{1} & \frac{1}{120} & -\frac{1}{360} \\ \hline
\frac{1}{2} & 4N_{1} & \frac{1}{40} & -\frac{11}{720} \\ \hline
1 & N_{1} & -\frac{3}{20} & -\frac{31}{180}%
\end{array}
\label{Coeffs}
\end{equation}%
and where 
\begin{eqnarray*}
H_{\mu \nu }^{(1)} &=&2R_{~;\mu \nu }-2g_{\mu \nu }\square R-\frac{1}{2}%
g_{\mu \nu }R^{2}+2RR_{\mu \nu } \\
H_{\mu \nu }^{(3)} &=&R_{\mu }^{\phantom{\mu}\rho }R_{\rho \nu }-\frac{2}{3}%
RR_{\mu \nu }-\frac{1}{2}R_{\rho \sigma }R^{\rho \sigma }g_{\mu \nu }+\frac{1%
}{4}R^{2}g_{\mu \nu }
\end{eqnarray*}%
Relating the parameters of the gravity theory in the bulk and those of the
CFT on the boundary gives 
\begin{equation}
\frac{1}{G}=\frac{2N^{2}}{\pi \ell ^{3}}  \label{Pararel}
\end{equation}%
The expression for the energy of the field theory is given by 
\begin{equation}
\mathcal{E}=\sum_{s=0,\frac{1}{2},1}n^{s}\int_{\Sigma }d^{3}x~\sqrt{\sigma }%
N_{lp}\left\langle \hat{T}_{ab}^{s}\right\rangle \xi ^{a}u^{b}
\label{EfromT4d}
\end{equation}%
where $\xi ^{a}=[1,0,0,0]$ and $u_{a}=\left[ {\ \left( -g^{tt}\right) ^{-1/2}%
},0,0,0\right] $, $N_{lp}$ is the lapse function, and the boundary metric is
decomposed as: 
\begin{equation}
ds^{2}=-N_{lp}^{2}dt^{2}+\sigma _{ab}\left( dx^{a}+N^{a}dt\right) \left(
dx^{b}+N^{b}dt\right)
\end{equation}%
Using (\ref{Tab4d}) in (\ref{EfromT4d}), the energy can be calculated to be 
\begin{equation}
\mathcal{E}=\sum_{s=0,\frac{1}{2},1}n^{s}\frac{4\pi ^{2}(\alpha ^{s}-3\beta
^{s})}{\ell }  \label{EcalcEH}
\end{equation}%
(where $\theta \in \left[ 0,\pi \right] $, $\phi \in \left[ 0,2\pi \right] $
and $\psi \in \left[ 0,4\pi \right] $). Note that the conserved mass is
given by (\ref{Mass5d}), or using (\ref{Pararel}) 
\begin{equation}
\mathfrak{M}=\frac{(3\ell ^{4}-4a^{4})N^{2}}{16\ell ^{5}p}  \label{MconsEH2}
\end{equation}
Now, taking (\ref{EcalcEH}) with (\ref{Coeffs}), and with $N_{1}=N^2$ (and
dividing by ($16\pi ^{2}$)) gives 
\begin{equation}
\mathcal{E}_{calc}=\frac{3N^{2}}{16p\ell }  \label{CasE}
\end{equation}%
which agrees with the Casimir energy from (\ref{MconsEH2}) with $a=0$. Note
also that a straightforward computation of the total energy using the No\"{e}%
ther charge approach \cite{GarfMann,Lorenzo} yields the difference $ 
\mathfrak{M}-\mathcal{E}_{calc}$, as expected.

The stress-tensor components from the conformal field theory can be written 
\begin{eqnarray}
\hat{T}_{tt} &=&-\frac{3N^{2}}{32\pi ^{2}\ell ^{4}}  \notag \\
\hat{T}_{\psi \psi } &=&-\frac{N^{2}}{128\pi ^{2}\ell ^{2}}  \notag \\
\hat{T}_{\psi \phi } &=&-\frac{N^{2}\cos (\theta )}{128\pi ^{2}\ell ^{2}}
\label{Tcft} \\
\hat{T}_{\theta \theta } &=&-\frac{N^{2}}{128\pi ^{2}\ell ^{2}}  \notag \\
\hat{T}_{\phi \phi } &=&-\frac{N^{2}}{128\pi ^{2}\ell ^{2}}  \notag
\end{eqnarray}%
These components match the (\ref{Tgravity}) components of $\mathcal{O}\left( 
{\textstyle\frac{1}{r^{2}}}\right) $ with $r=\ell $, $a=0$ and using the
transformation (\ref{Pararel}).

We see that the energy (\ref{Mass5d}) of the EH soliton is lower than that
of the AdS$_{5}/\Gamma $ orbifold, and is in fact negative once the
condition (\ref{EHmatch}) is taken into account. For any given integer $%
p\geq 3$ we have 
\begin{equation}
\mathcal{E}_{\text{EH-soliton}}=-\frac{(p^{4}-8p^{2}+4)N^{2}}{64\ell p}
\label{Energy}
\end{equation}%
which is bounded from below. In a previous paper \cite{EHshort}, we
conjectured that the EH soliton is the state of lowest energy in its
asymptotic class in both 5D Einstein gravity with negative cosmological
constant and in type IIB supergravity in 10 dimensions. Indeed, the AdS/CFT
correspondence (along with the expected stability of the gauge theory)
suggests that any metric solving the 5D Einstein equations that has the same
boundary conditions as the EH soliton will have a greater energy. This
situation is analogous to that for the AdS soliton \cite{HoroMyers}.

We demonstrate the our conjecture holds at least perturbatively. We wish to
construct the energy density $H$ directly to second order in the
fluctuations of the metric, given by $h_{\mu \nu }$, 
\begin{equation}
g_{\mu \nu }=\bar{g}_{\mu \nu }+h_{\mu \nu }  \label{pertmet}
\end{equation}%
where $\bar{g}_{\mu \nu }$\ is the EH solition (\ref{mtrcEHdS5}) and the
perturbation obeys the falloff conditions 
\begin{equation*}
h_{\mu \nu }=\mathcal{O}\left( r^{-2}\right) \text{ \ \ \ \ \ \ \ }h_{\mu r}=%
\mathcal{O}\left( r^{-4}\right) \text{ \ \ \ \ \ \ \ }h_{rr}=\mathcal{O}%
\left( r^{-6}\right) \text{ \ \ \ \ \ \ }\mu ,\nu \neq r
\end{equation*}%
Employing the method of Abbot and Deser \cite{AbbotDeser} (see also \cite%
{HoroMyers}), the Hamiltonian $H$ on a time-symmetric slice to second order
in the perturbation $h_{ij}$ ($i,j\neq t$) is 
\begin{equation}
\mathcal{H}=\bar{N}\left[ \frac{1}{\sqrt{\bar{g}}}p^{ij}p_{ij}+\sqrt{\bar{g}}%
\left( \frac{1}{4}\left( \bar{D}_{k}h_{ij}\right) ^{2}+\frac{1}{2}{}^{(4)}%
\bar{R}^{ijkl}h_{il}h_{jk}-\frac{1}{2}{}^{(4)}\bar{R}^{ij}h_{ik}h_{j}^{%
\phantom{j}k}\right) \right]
\end{equation}%
where to arrive at this equation, the gauge conditions 
\begin{equation}
p_{\phantom{i}}^{i}=0=\bar{D}^{i}h_{ij}
\end{equation}%
are imposed, with $p_{ij}$ the conjugate momentum. We further impose 
\begin{equation}
h_{\phantom{i}i}^{i}=0=\bar{D}_{i}p^{ij}
\end{equation}%
as they are required to satisfy the constraint equations to linear order.

It is easily seen that the momenta make a positive contribution to the
energy density. We therefore choose intial conditions such that $p^{ij}=0$
in order to minimize the energy of the spacetime with metric $g_{\mu \nu }$.
Thus we need only calculate the gradient energy density (also positive) and
the potential energy density, given by the terms 
\begin{eqnarray}
U &=&\frac{1}{2}~{}^{(4)}\bar{R}^{ijkl}h_{il}h_{jk}-\frac{1}{2}~{}^{(4)}\bar{%
R}^{ij}h_{ik}h_{j}^{\phantom{j}k} \\
&=&\frac{1}{2\ell ^{2}}V^{a}U_{ab}V^{b}
\end{eqnarray}
where 
\begin{equation}
V^a = \left[ h_{\psi \psi}, h_{\theta \theta}, h_{\phi \phi}, h_{\psi r},
h_{\psi \theta}, h_{\psi \phi}, h_{r\theta}, h_{r \phi}, h_{\theta \phi} %
\right]
\end{equation}
and so the problem reduces to that of finding the eigenvalues of $U_{ab}$.
Note that this vector only has 9 components - the $h_{rr}$ component in the
matrix $U_{ab}$ will be removed below from the traceless condition that must
be applied.

In the EH case, the curvature of the spatial slice in the orthonormal frame
is given by the components 
\begin{eqnarray}
{}^{(4)}R_{\psi r\psi r} &=&\frac{-r^{6}+3r^{2}a^{4}+4a^{4}\ell ^{2}}{\ell
^{2}r^{6}} \\
{}^{(4)}R_{\theta \phi \theta \phi } &=&\frac{4a^{4}\ell
^{2}-r^{6}+r^{2}a^{4}}{\ell ^{2}r^{6}} \\
{}^{(4)}R_{\psi \theta \psi \theta } &=&\frac{-r^{6}-r^{2}a^{4}-2a^{4}\ell
^{2}}{\ell ^{2}r^{6}}={}^{(4)}R_{\psi \phi \psi \phi }={}^{(4)}R_{r\theta
r\theta }={}^{(4)}R_{r\phi r\phi } \\
{}^{(4)}R_{\psi r\theta \phi } &=&-\frac{4a^{4}\sqrt{r^{2}+\ell ^{2}}}{%
r^{6}\ell } \\
{}^{(4)}R_{\psi \theta r\phi } &=&-\frac{2a^{4}\sqrt{r^{2}+\ell ^{2}}}{%
r^{2}\ell }=-{}^{(4)}R_{\psi \phi r\theta }
\end{eqnarray}%
Because of the form of these components, the matrix $U_{ab}$ in the
potential energy density above must be written as a 9$\times $9 matrix. This
matrix is block-diagonal, with the ``diagonal'' components (those components
that can be written as coefficients of $h_{ii}^{2}$) being the upper
diagonal 4$\times $4 matrix, and the ``off-diagonal'' components (those that
can be written as coefficients of $h_{ij}h_{kl}$, where $i\neq j$ and $k\neq
l$) being the lower 5$\times $5 matrix.

Dealing with the diagonal components first, here we can apply the traceless
condition $h_{rr}=-h_{\psi \psi }-h_{\theta \theta }-h_{\phi \phi }$ to
reduce this to a 3$\times $3 matrix, giving the symmetric matrix 
\begin{equation}
U_{ab(diag)}=\left[ 
\begin{array}{ccc}
D_{11} & D_{12} & D_{13} \\ 
D_{21} & D_{22} & D_{23} \\ 
D_{31} & D_{32} & D_{33}%
\end{array}%
\right]
\end{equation}%
where the components are 
\begin{eqnarray}
D_{11} &=&\frac{4(r^{6}+r^{2}a^{4}+2a^{4}\ell ^{2})}{r^{6}}=2D_{12}=2D_{13}
\\
D_{22} &=&\frac{2(2r^{6}-r^{2}a^{4}-2a^{4}\ell ^{2})}{r^{6}}=D_{33} \\
D_{23} &=&\frac{2(r^{6}-4a^{4}\ell ^{2}-2r^{2}a^{4})}{r^{6}}
\end{eqnarray}%
The eigenvalues of this matrix are easily found to be 
\begin{eqnarray}
\lambda _{0} &=&\frac{2(r^{6}+r^{2}a^{4}+2a^{4}\ell ^{2})}{r^{6}} \\
\lambda _{\pm } &=&\frac{1}{r^{6}}\Big[5r^{6}-r^{2}a^{4}-2a^{4}\ell ^{2} 
\notag \\
&&\pm \sqrt{9r^{12}+6r^{8}a^{4}+12a^{4}r^{6}\ell
^{2}+33r^{4}a^{8}+132r^{2}a^{8}\ell ^{2}+132a^{8}\ell ^{4}}\Big]
\end{eqnarray}%
We can plot these eigenvalues as a function of $x=a^{4}/r^{4}$, taking into
account the regularity condition $a=\ell \sqrt{p^{2}/4-1}$. One can easily
see from such plots that $\lambda _{0}$ and $\lambda _{+}$ are positive in
the range of interest $0\leq x\leq 1$, and hence correspond to stable
fluctuations. However, a plot of $\lambda _{-}$ shows that this eigenvalue
is negative for $x\gtrsim 0.25$ for $p=3$,\ and becomes negative for $%
x\gtrsim 0.5$ when $p\rightarrow \infty $. Thus there is a region where the
potential energy density becomes negative; the associated eigenvector is
given by 
\begin{eqnarray}
V_{diag}^{a} &=&[v_{m},1,1] \\
v_{m} &=&\frac{1}{2(r^{6}+r^{2}a^{4}+2a^{4}\ell ^{2})}\Big[%
5r^{2}a^{4}-r^{6}+10a^{4}\ell ^{2} \\
&&-\sqrt{9r^{12}+6r^{8}a^{4}+12a^{4}r^{6}\ell
^{2}+33r^{4}a^{8}+132r^{2}a^{8}\ell ^{2}+132a^{8}\ell ^{4}}\Big]  \notag
\end{eqnarray}

Writing the perturbation as $h_{ik}=A\left( r\right) \tilde{h}_{ik}$, where $
A\left( r\right) $ is a profile function maximized at $r=a$\ and $\tilde{h}%
_{ik}$ is the eigenvector associated with the negative eigenvalue, we find
that the negative potential energy $U$ is not outweighed by a simple
estimation of the gradient energy density (given by dividing the maximum of
the profile function by the proper distance over which $U$ is negative).
This situation - quite unlike that for the AdS soliton \cite{HoroMyers} -
forces us to consider a full expansion of the gradient energy term. Using
this, we find that the full gradient energy density always outweighs the
potential energy density, indicating that the diagonal component of the
potential energy density is perturbatively stable for all values of $p$.

The situation proceeds similarly for the off-diagonal components of $U$.
Here, we get the 5$\times $5 matrix 
\begin{equation}
U_{ab(off-diag)}=\left[ 
\begin{array}{cccccc}
E_{11} & 0 & 0 & 0 & 0 & 0 \\ 
0 & E_{22} & 0 & 0 & E_{25} & 0 \\ 
0 & 0 & E_{33} & E_{34} & 0 & 0 \\ 
0 & 0 & E_{43} & E_{44} & 0 & 0 \\ 
0 & E_{52} & 0 & 0 & E_{55} & 0 \\ 
0 & 0 & 0 & 0 & 0 & E_{66}%
\end{array}%
\right]
\end{equation}%
(where the order of components is $[h_{\psi r},h_{\psi \theta },h_{\psi \phi
},h_{r,\theta },h_{r,\phi },h_{\theta \phi }]$). The components of $
U_{ab(off-diag)}$ are 
\begin{eqnarray}
E_{11} &=&\frac{4(r^{6}+r^{2}a^{4}+2a^{4}\ell ^{2})}{r^{6}}=E_{66} \\
E_{22} &=&\frac{2(2r^{6}-r^{2}a^{4}-2a^{4}\ell ^{2})}{r^{6}}%
=E_{33}=E_{44}=E_{55} \\
E_{25} &=&\frac{12a^{4}\ell (r^{2}+\ell ^{2})}{r^{6}\sqrt{r^{2}+\ell ^{2}}}%
=E_{52}=-E_{34}=-E_{43}
\end{eqnarray}

The eigenvalues of this matrix are 
\begin{eqnarray}
\lambda _{0} &=&\frac{4(r^{6}+r^{2}a^{4}+2a^{4}\ell ^{2})}{r^{6}} \\
\lambda _{\pm } &=&\frac{2(2r^{6}-r^{2}a^{4})}{r^{6}}-\frac{4a^{4}\ell ^{2}}{%
r^{6}}\pm \frac{12a^{4}\ell \sqrt{r^{2}+\ell ^{2}}}{r^{6}}
\end{eqnarray}%
Here again $\lambda _{0}$ and $\lambda _{+}$ are positive in the range $%
0\leq x\leq 1$, but the $\lambda _{-}$ eigenvalue is negative. Unlike the
diagonal case, the negative eigenvalue is only negative for $3\leq p\leq 14$
(recall $p$ is an integer).

With the same perturbation $h_{ik}=A\left( r\right) \tilde{h}_{ik}$ as in
the diagonal case, we find that the full expansion of the gradient term is
again required, but that again, the full gradient energy density does always
outweigh this part of the potential energy density.

This indicates that the EH soliton is perturbatively stable for all values
of $p$ relative to all other metrics with the same boundary conditions.
Together with the expected stability of the (nonsupersymmetric) gauge
theory, the AdS/CFT correspondence suggests that the EH soliton is a
solution of minimal energy, and that any other solution with these boundary
conditions will have an energy greater than this. We are therefore led to
make conjectures for these spacetimes similar to those for the AdS soliton:

Conjecture 1: All solutions to ten-dimensional IIB supergravity satisfying ( %
\ref{pertmet}), with $\bar{g}_{\mu \nu }$\ given by the metric (\ref%
{mtrcEHdS5}), will have energy greater than this unless $h_{\mu \nu }=0$.

Conjecture 2: All solutions to Einstein's equation in five dimensions with
negative cosmological constant satisfying (\ref{pertmet}), with $\bar{g}
_{\mu \nu }$\ given by the metric (\ref{mtrcEHdS5}), will have energy
greater than this unless $h_{\mu \nu }=0$.

Conjecture 3: Any nonsingular Riemannian 4-manifold with negative Ricci
scalar $R=-\frac{12}{\ell ^{2}}$ satisfying (\ref{pertmet}), with $\bar{g}%
_{\mu \nu }$\ given by the metric (\ref{mtrcEHdS5}), will have energy
greater than this unless $h_{\mu \nu }=0$.

The motivation for these conjectures is the same as that for the AdS
soliton. If spacetimes in 10 dimensions that are not direct products with $%
S^{5}$ have higher energy, and if the additional supergravity fields
contribute positive energy, then conjecture 1 reduces to conjecture 2.
Similarly, if there exists a moment of time symmetry (i.e. there is a
surface with zero extrinsic curvature), then conjecture 2 reduces to
conjecture 3. It would be interesting to see if such conjectures could be
proven.

\subsection{dS/CFT}

The proposed dS/CFT correspondence arises from the AdS/CFT by analogy, and
works in similar fashion, though with difficulties that don't arise in the
AdS/CFT -- namely, that going to future/past infinity in an asymptotically
dS spacetime means the timelike killing vector $\partial /\partial t$
becomes spacelike as one is now outside the cosmological horizon. The
conserved charge associated with this Killing vector is interpreted as the
conserved mass. With this definition, Balasubramanian \textit{et. al.} \cite%
{Bala} were led to posit the conjecture \textit{any asymptotically dS
spacetime with mass greater than dS has a cosmological singularity}, which
was referred to as the maximal mass conjecture in \cite{CGML}. There, the
Taub-NUT-dS (TNdS) spacetime was shown to be a counter-example to this
conjecture, though there have been arguments as to whether it is a true
counter-example due to the existence of closed timelike curves (CTC's) in
the TNdS spacetime.

We find for the EH-dS metrics in five dimensions that 
\begin{equation*}
\mathfrak{M}=\frac{\pi (3\ell ^{4}-4a^{4})}{32pG\ell ^{2}}  \label{dsmass}
\end{equation*}%
in agreement with the maximal mass conjecture, since whenever $a<\ell$ the
regularity condition implies that 
\begin{equation*}
\mathfrak{M}=\frac{3\ell^2\pi}{128 G}<\mathfrak{M}_{dS} = \frac{3\ell^2 \pi}{%
32}
\end{equation*}%
and whenever $a\geq \ell$ it is clear that $\mathfrak{M}<0$.

The curious thing about the metric (\ref{mtrcEHdS5}) for the EH-dS case is
that, after applying the dS/CFT counter-term approach, we calculate the same
quantities (\ref{Mass5d}), (\ref{Itot}) and (\ref{Stot}) for the conserved
mass, action and entropy of the EH-AdS metric. Note, however that this is a
formal agreement. The regularity condition in the AdS case implies that $p>3$
(and that $p$ is even if fermions are present in the boundary CFT), whereas
in the dS case this condition implies when $a<\ell$ that either $p=1$
(yielding the mass (\ref{dsmass}) or when $a>\ell$ that 
\begin{equation*}
\mathfrak{M}=-(p^4+8p^2+4)\frac{\ell^2\pi}{128 p G}
\end{equation*}%
which differs from the AdS mass given by (\ref{Energy})). In the dS case
there is a horizon, where $r_{+}=\ell>a$, and so the action becomes 
\begin{equation}
I_{tot(dS)}=\frac{\beta \pi (4a^{4}-5\ell ^{4})}{32\ell ^{2}}  \label{Itotds}
\end{equation}%
Through the Gibbs--Duhem relation $S=\beta \mathfrak{M}-I$, the entropy is 
\begin{equation}
S_{tot(dS)}=\frac{\beta \pi (\ell^{4}-a^{4})}{4\ell^{2}}  \label{Stotds}
\end{equation}%
where the regularity condition has not been applied.

The mass (\ref{Mass5d}) and entropy (\ref{Stotds}) can be shown to satisfy
the first law $dS=\beta d\mathfrak{M}$, and the differentiation to be taken
w.r.t. $a$, where 
\begin{equation*}
\beta =\frac{2\pi \ell ^{3}}{\sqrt{\ell ^{4}-a^{4}}}
\end{equation*}%
is obtained from the surface gravity at the horizon. The specific heat for
the EHdS case can also be calculated, giving 
\begin{eqnarray}
C_{dS} &=&-\beta \partial _{\beta }S=-\beta \frac{\partial a}{\partial \beta 
}\frac{\partial S}{\partial a}  \notag \\
&=&\frac{\pi ^{2}\ell \sqrt{\ell ^{4}-a^{4}}}{2}  \label{Ctot}
\end{eqnarray}%
Substituting in $\beta $ into (\ref{Stotds}) for the EHdS case gives the
same formula as the specific heat - (\ref{Ctot}). Thus, the solution can be
considered thermodynamically stable for equation (\ref{Ctot}) greater than
zero, i.e. for $\ell >a$. Since for $a>\ell $, the solution would be
imaginary, this means that for all real entropies, the solution is
thermodynamically stable.

Applying the regularity condition $a^2=\frac{3}{4}\ell^2$ yields 
\begin{eqnarray*}
\mathfrak{M} & = & \frac{a^2\pi}{32 G} \\
\beta &=& \frac{48\pi a}{\sqrt{21}} \\
S & = & \frac{\pi ^{2}a^{2}\sqrt{21}}{9}
\end{eqnarray*}%
for the various thermodynamic quantities. Note that the first law is no
longer satisfied since the thermodynamic variable $a$ has been constrained
in terms of the independent parameter $\ell$.

\section{Discussion}

We have presented a new class of soliton solutions to the higher-dimensional
Einstein equations with cosmological constant. In the limit that the
cosmological constant vanishes they are a direct product of a
higher-dimensional version of the Eguchi-Hanson metric with time. For this
reason we have called such solutions Eguchi-Hanson solitons.

EH solitons have a variety of interesting properties. For negative
cosmological constant they provide an explicit example of solutions that
asymptotically approach AdS$_{d+1}/\Gamma $ but have lower energy, with $%
\Gamma=Z_p$. The ground state energy density of the strongly coupled gauge
theory on the boundary on $S^{d-1}/\Gamma $ is then even smaller than on $%
S^{d-1}$. For $d=4$ we provided evidence that perturbatively the AdS
EH-soliton is the state of lowest energy in its asymptotic class. We
conjecture that this holds non-perturbatively (for all dimensions) as well,
a proof (or refutation) of which remains an interesting subject for future
study.

In the dS case all such solutions have a cosmological horizon. If the
constant of integration $a$ is large enough then the spacetime has closed
timelike curves. For all values of $a$ these solutions satisfy the maximal
mass conjecture, ie they all have mass less than that of pure de Sitter
spacetime with the same asymptotics.

Further analysis of these metrics, including their possible phase-transition
properties, their Kaluza-Klein reduction, and their role in string theory
remain interesting issues to explore in the future.

\bigskip \noindent \textbf{Acknowledgements}

We are grateful to Gary Horowitz, Ted Jacobson, Juan Maldecena, Rob Myers,
Kristin Schleich, Don Witt and Eric Woolgar for helpful discussions and
correspondence. This work was supported in part by the Natural Sciences and
Engineering Research Council of Canada.

\end{document}